\documentclass[aps,pre,preprint,showpacs,superscriptaddress]{revtex4-1}

\usepackage{amsmath}
\usepackage{amssymb}
\usepackage{hyperref}
\usepackage{graphicx}
\usepackage{subfigure}
\usepackage{overpic}
\usepackage{tikz}
\usepackage{comment}
\usepackage{siunitx}

\usetikzlibrary{calc}
\tikzset{egrid/.style={draw,help lines}}
\tikzset{mgrid/.style={draw,help lines,dashed}}
\tikzset{epoint/.style={draw,circle,red,inner sep=2pt,fill}}
\tikzset{mpoint/.style={draw,circle,blue,inner sep=2pt,fill}}

\renewcommand{\vec}{\mathbf}
\newcommand{\eps}{\varepsilon}

\begin{document}

\title{Generation of vector beams with liquid crystal disclination lines}

\author{Miha \v Can\v cula}
\author{Miha Ravnik}
\affiliation{Faculty of Mathematics and Physics, University of Ljubljana, Slovenia}

\author{Slobodan \v Zumer}
\affiliation{Faculty of Mathematics and Physics, University of Ljubljana, Slovenia}
\affiliation{Jo\v zef Stefan Institute, Ljubljana, Slovenia}

\date{\today}

\begin{abstract}

We report that guiding light beams, ranging from continuous beams to femtosecond pulses, along liquid crystal defect lines can transform them into vector beams with various polarization profiles. 
Using Finite Difference Time Domain numerical solving of Maxwell equations, we confirm that the defect in the orientational order of the liquid crystal induces a defect in the light field with twice the winding number of the liquid crystal defect, coupling the topological invariants of both fields. 
For example, it is possible to transform uniformly-polarized light into light with a radial polarization profile.
Our approach also correctly yields a zero-intensity region near the defect core, which is always present in areas of discontinuous light polarization or phase. 
Using circularly polarized incident light, we show that defects with non-integer winding numbers can be obtained, where topological constants are preserved by phase vortices, 
demonstrating coupling between the light's spin, orbital angular momentum and polarization profile. 
Further, we find an ultrafast femtosecond laser pulse travelling along a defect line splits into multiple intensity regions, again depending on the defect's winding number, allowing applications in beam steering and filtering. 
Finally, our approach describing generation of complex optical fields via coupling with topological defect lines in optically birefringent nematic fluids can be easily extended to high-intensity beams that affect nematic ordering.

\end{abstract}


\maketitle

\section{Introduction}

Vector light beams are distinct optical fields characterized by complex spatial modulation of both intensity profile and the polarization \cite{Hall:96}. Theoretically, they are solutions of the vector paraxial wave equation with a non-uniform polarization profile \cite{Zhan:04,Tovar:98}. Especially interesting are cylindrical vector beams ({\small CVB}) with an axially symmetric intensity profile, which can be focused to a smaller spot size \cite{radial-focus} and are useful in optical particle trapping \cite{Zhan:radial-trap,radial-trapping-forces}. Furthermore, cylindrical vector beams have  applications in optical microscopy \cite{Youngworth:cvb-focusing-na} and in laser cutting \cite{radial-cutting}. 
They can be generated from linearly polarized scalar beams, such as those emmited by most lasers, by using conical Brewster prisms \cite{kozawa-sato-laser}, few-mode fibers \cite{cvb-fibers}, or -- today most commonly -- by phase modulators \cite{polarization-converters-axial}. 

Liquid crystals are central materials in modern display optics and photonics due to their birefringence and their susceptibility to external control by electric, magnetic and optical fields. The birefringence in nematics stems from the orientational order of molecules which align along a distinct direction, called the director, corresponding to the optical axis \cite{degennes,kleman}. The optical axis can spatially vary over mutiple spatial scales, from \SI{10}{\nano\meter} to \SI{10}{\micro\meter}, notably also including the visible wavelength range, which makes nematics attractive for complex optic and photonic devices \cite{coles-morris, Assanto:solitons}. Of special interest for photonics are topological defects in the orientational order,  where the director field is discontinuous, and the material becomes locally optically isotropic \cite{colloquium}. These nematic defects can be in the form of points or lines, and each defect is characterized by distinct spatial variation of optical axis around the defect. 
Line defects, also called disclinations, are characterized by the topological invariant called winding number (also known as topological charge or strength), which specifies how many turns the optical axis (director) makes when following a loop around the disclination. 
The director is a headless vector, so opposite directions are equivalent, allowing for disclination lines with half-integer and integer winding numbers. 
Disclinations carry a high free energy cost, so their presence has to be enforced by topological constrains, boundary conditions or strong external fields \cite{kleman,Araki:confined,opali}. 

There is a fundamental connection between the liquid crystal director and the light field if the two get coupled, for example by shining light field on a structure in the nematic director field.
It has been shown that nematic disclinations can induce singularities in the light field \cite{brasselet-droplet}, and sufficiently strong light fields can imprint defects in liquid crystals \cite{brasselet-reordering,porenta}. 
Generation of scalar and vector vortex beams on disclinations lines in nematic films was also recently experimentally demonstrated \cite{brasselet-film}. 
Manipulation of vector beam polarization patterns can be achieved by using a liquid crystal device called a ``q-plate'' \cite{Marucci:06}. 
It has been used for imprinting polarization singularities into light beams, with either zero-intensity regions or points with circular polarization at the core \cite{Cardano:13}. 
The q-plate consists of a liquid crystal disclination line with a certain winding number and a fixed length, where the director structure is enforced by treating the surface. 
A disclination line with winding number $s$ can transform a linearly polarized scalar beam into a beam with a polarization defect with winding number $2s$, as can be shown using a single-photon state formalism \cite{Nagali:09} or using the Jones formalism \cite{Cardano:12}. 
A light beam carries spin, manifesting itself in the classical picture as cylindrical polarization of light, as well as orbital angular momentum ({\small OAM}) if a phase vortex is present at the beam axis\cite{humblet:43}. 
Transfer of angular momentum between spin and orbital degrees of freedom can be achieved in certain photonic devices \cite{brasselet-film,Zhao:spin-orbital}, including the q-plate \cite{Marucci:06}. 
Angular momentum transfer can also be calculated analytically in the paraxial approximation and assuming low material birefringence \cite{Karimi:09}, but moving beyond these approximations requires numerical methods. 

Our interest will be in modelling the coupling between the light field and nematic disclinations, and there exist multiple numerical methods for modelling the propagation of light through anistropic birefringent media. 
Jones calculus describes light with two transversal components of the electric field enforcing a fixed direction of propagation and neglecting diffraction.
It is suitable for uniform media or media with spatial variations of refractive indices on a scale much larger than the wavelength of light. 
For more complex anisotropic media, the Berreman method can be used \cite{berreman,stallinga-berreman}, which calculates the transversal components of both electric and magnetic fields and can describe more rapid spatial variations. 
However, these simplified descriptions fail where the diffaction and refraction cause the propagation of light in $x$ and $y$ direction to be significant, such as in the presence of discontinuities of the permittivity tensor \cite{metode-kriezis,hwang-rey}. 
In these cases, a method that considers all six components of light fields must be used.
The Finite Difference Time Domain (FDTD) method has proven to be a powerful tool for modelling light propagation through arbitrarily complex optically anisotropic media, computing the time evolution of the electric and magnetic field by explicitely solving Maxwell's equations\cite{Johnson:Meep,taflove}.
Indeed, along with the Finite Element Method \cite{Neyts:fem}, it is often used especially in the study of liquid crystal optics and photonics \cite{Kriezis99,Ogawa:13,Matsui:fdtd}.

In this paper, we present numerical modelling of the flow-of-light along liquid crystal disclination lines, showing that this approach allows for controllable design of both polarization and intensity of cylindrical vector light beams.
Using a custom developed FDTD-based numerical approach, we find that a $+1/2$ disclination line can transform a scalar beam into a beam with radial or azimuthal polarization, depending on the polarization of incident light. 
By employing nematic disclination lines with different winding numbers ($-1,-1/2,+1/2,+1$), we find vector beams with further polarization profiles. 
Our results confirm the defect winding numbers predicted by the Jones formalism, while additionally demonstrating the necessary zero-intensity region at the defect core. 
Owing to the sensitivity of liquid crystals, the polarization conversion can be tuned by direct system parameters, including temperature, birefringence, and the length of the disclination.
We show that light polarization always forms a defect with twice the winding number of the liquid crystal disclination and provide a simple analytical derivation that explains this relation. 
Unlike derivations based on photon states or the Jones formalism, our FDTD method correctly shows a zero-intensity region, which ensures that electric and magnetic fields are continuous near the defect core. 
By using circularly polarized incident light, we observe that it is possible to induce polarization defects with {\it half-integer} winding numbers, which  is seemingly counter-intuitive and should have been prevented by the vector symmetry of the light polarization. 
However, this incompatibility is resolved by a phase singularity at the beam axis, demonstrating coupling between spin, orbital angular momentum and polarization.
In addition to continuous light beams, we present light field modulation of femtosecond laser pulses by nematic disclinations.
We assume low light intensity and neglect any effects that light has on the orientation or temperature of the liquid crystal. 
The light pulses undergo similar transformation of their polarization state, but interestingly also split into multiple intensity regions. 
The number of these intensity regions is found to be equal to $2|s|$, where $s$ is the disclination line's winding number. 

\section{Methods}
There is a mutual interaction between a nematic liquid crystal and light.
Nematic birefringence affects the polarization of light, while the liquid crystal molecules generally tend to orient parallel to the electric field. 
However, in this paper, we assume that light fields are too weak to reorient the nematic, and only model the flow of light through a fixed nematic director field. 
We develop and implement a custom written finite-difference time-domain ({\small FDTD}) approach, which allows for modelling of the flow-of-light in general optically anisotropic materials, with notably spatially varying optical axis, such as liquid crystals. The method computes the values of all six components of electromagnetic fields at every time step by following the time derivatives of two dynamical Maxwell's equation\cite{taflove}. 
Using dimensionless quantities ($c = \mu_0 = \eps_0 = 1$) and assuming no free charges and currents in the material, these equations are
\begin{align}
\label{eq:maxwell-2}
 \frac{\partial \vec H}{\partial t} = - \nabla \times \vec E, \qquad \frac{\partial \vec E}{\partial t} = \eps^{-1}\left( \nabla \times \vec H \right),
\end{align}
where $\vec E$ and $\vec H$ are electric and magnetic field, respectively. 
The two equations contain one material field, i.~e.~the dielectric permittivity tensor $\eps$, and it is the design and spatial variation of $\eps=\eps(\vec r)$ tensor which is at the center of this paper, driving the polarization and intensity modulation of the vector light beams. 

\begin{figure}[ht]
\centering
\includegraphics[width=.8\textwidth]{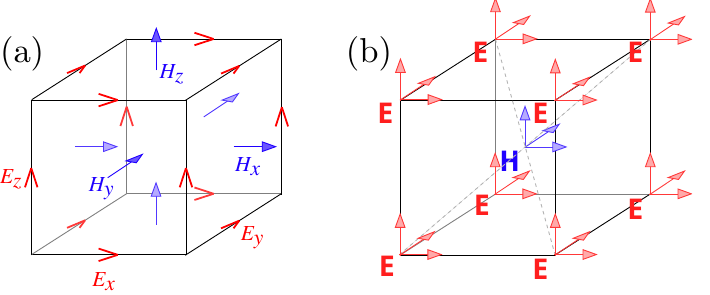}
\caption{(Color online)
  (a) The Yee lattice, where each field component is computed at a different site.
  (b) The lattice we used, with electric field known at cube vertices and magnetic field at cube centers.
  In both cases, the electric and magnetic fields are computed at different times. 
}
\label{fig:lattice}
\end{figure}

When using Eqs.~\ref{eq:maxwell-2} on a discrete lattice in the FDTD approach, it is possible to improve accuracy without sacrificing performance by defining field components at different points in space and time. The standard apporach is to use the Yee lattice\cite{yee,Johnson:Meep} (Figure~\ref{fig:lattice}a), where each field component is known at a different site. 
This way, all space-derivatives are of the central-difference type, giving the method second-order accuracy. 
However, this is true only if the components of the electric fields are uncoupled, i. e. if the permittivity tensor $\eps$ is diagonal, which can be achieved in a {\it uniform} birefringent material by choosing a suitable frame of reference. 
In liquid crystals, where the dielectric tensor and birefringent axes inherently vary in space, such a choice cannot be made, so the inherent symmetry of the Yee lattice is lost. 
It is possible to use the Yee lattice with arbitrary permittivity tensors, but this causes the loss of the central-difference property, reducing the accuracy and giving rise rise to numerical stability concerns\cite{Werner:2007:stabilnost}. 
To alleviate these shortcomings, we use our own lattice, where notably, all three components of each of the two fields (i.e. electric and magnetic) are computed at the same point (see Figure~\ref{fig:lattice}b). 
The introduced lattice has the same second-order accuracy as the Yee lattice and is unconditionally stable\cite{Werner:2007:stabilnost}, at the cost of slightly decreased computational performance. 
To ensure all differences are central, each space-derivative has to be computed as an average of four finite differences, as opposed to one as with the Yee lattice. 
This larger number of operations is offset by adapting the method to support parallel computing on CPU or GPU, resulting in major performance gains. 
For example, a fully 3D simulation run with 2000 time steps (corresponding to cca \SI{300}{\femto\second}) on $\sim 5\times 10^7$ lattice points only takes a few minutes on a single Nvidia GTX Titan {\small GPU}.
External light sources are modelled by partitioning the simulation cell with a boundary and empty space around the cell with an absorbing perfectly-matched layer ({\small PML}) \cite{berenger}.
By setting the absorbing layer's thickness to 30 simulation points, reflection is reduced by five orders of magnitude, as commonly seen in {\small FDTD} modelling \cite{taflove,Johnson:Meep}. 

The liquid crystalline birefringent profile is introduced into the FDTD simulations via a spatially-varying permittivity tensor \cite{degennes}
\begin{align}
\label{eq:lc-permittivity}
 \eps_{ij} &= \overline\eps \delta_{ij} + \eps^\mathrm{a}_{ij} = \overline\eps\delta_{ij} + \frac{2}{3}\eps_\mathrm{a}^{\mathrm{mol}} Q_{ij}
\end{align}
where $\overline\eps$ is the average permittivity, $\eps_\mathrm{a}^{\mathrm{mol}} = \eps_{\parallel}^{\mathrm{mol}} - \eps_{\perp}^{\mathrm{mol}}$ is the molecular dielectric anisotropy, and $Q_{ij} = \frac{S}{2}(3n_i n_j - \delta_{ij})$ is the uniaxial nematic order parameter tensor. A fixed uniaxial nematic order parameter tensor is assumed here with the optical axis (corresponding to the extraordinary refractive index) equivalent to the nematic director $n_i$. In our calculations, we computed $Q_{ij}$, $\eps_{ij}$ and its inverse from the director $\vec n(\vec r)$ and the nematic degree of order $S$, which corresponded to the nematic disclinations. 
The director field around nematic disclination lines can be written as $\vec n(r, \phi, z) = \cos s \phi \; \hat{\vec e}_x + \sin s \phi \; \hat{\vec e}_y$
where $s$ is the disclination line's winding number and can be of integer or half-integer value. Energetically, the energy of disclinations per unit lenght are proportional to $s^2$ \cite{kleman}; therefore, typically, only disclinations with low winding numbers emerge in experiments\cite{degennes}, so we focused only on systems with $|s| < 2$. The defect cores are modelled by reducing the nematic degree of order $S$ smoothly to zero within the core radius of few $10~\rm{nm}$ \cite{degennes}, effectively, causing the material to become optically isotropic. Indeed, the core region being much smaller than the wavelength of light, we find that its exact structure has little effect on the propagation of light. Because $Q$, $\varepsilon$ and $\varepsilon^{-1}$ are all related symmetric tensors, only six values of $\varepsilon^{-1}$ are stored in claculations, saving on computer memory.

The {\small FDTD} method allows arbitrary incident light fields. 
We choose a scalar {\small TEM}$_{00}$ Gaussian beam with uniform polarization, either linear or circular. 
In the case of a pulse source, the field amplitude varies in time as
\begin{align}
 E_0(t) \propto \exp\left({-\frac{(t-t_0)^2}{2T^2}}\right)
\end{align}
where $T$ is the characteristic pulse length $T$. 
Because there is no light before $t=0$, effectively cutting of one tail of the Gaussian function, the peak-intensity time $t_0$ is chosen much larger than $T$, so that the full shape of the pulse was captured. 
A value of $t_0\approx5T$ appears sufficient to remove any artifacts and produce well-formed Gaussian-shaped pulses. 

Unless stated differently, numerical and modelling parameters in Table \ref{tab:parameters} are used, corresponding to standard nematic liquid crystal (e.g. 5CB)\cite{kolicniki} and visible light. Free-standing disclination lines are modelled without an external light-guiding structure, so the beams diverge while traversing the disclinations. However, the Rayleigh range in the medium -- a characteristic distance from the waist at which the cross section area of the beam is doubled -- is greater than the simulation cell size, ensuring that while beam divergence is noticeable, its effect on the polarization and phase profiles is negligible. 

\begin{table}[h]
 \begin{tabular}{|lrl|}
 \hline
 Ordinary refractive index & $n_\mathrm{o} = 1.54$ & \\
 Extraordinary refractive index & $n_\mathrm{e} = 1.69$ & \\
 Light wavelength & $\lambda = 480$ & nm\\
 Lattice resolution & 30 & nm\\
 Lattice size & $224 \times 224 \times 640$ & voxels \\
 & $6.7\times 6.7\times 19.2$ & \SI{}{\micro\meter} \\
 Beam waist & $w_0 = 1600$ & nm \\
 Pulse length & $T = 6$ & fs \\
 \hline
 \end{tabular}
 \caption{Used numerical parameters of the liquid crystal and light}
 \label{tab:parameters}
\end{table}

\newcommand{\zd}{\ensuremath{z_\mathrm{d}}}

\section{Results}

\subsection{Linear incident polarization}

\begin{figure*}[!ht]
 \includegraphics[width=\textwidth]{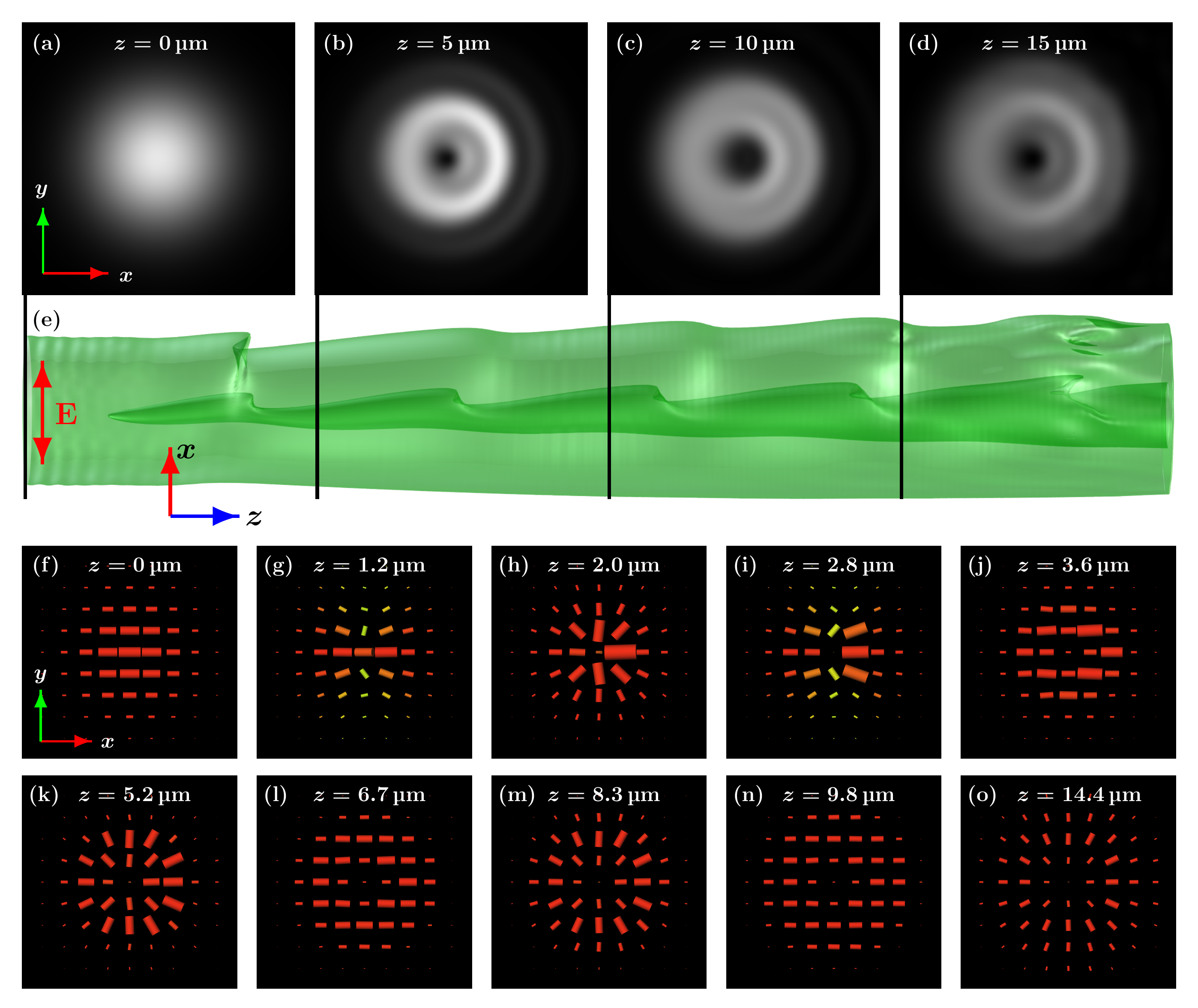}
 \caption{(Color online)
  Generation of radial vector light beam with $+1/2$ nematic disclination line. 
  (a-d) Light intensity profiles at different cross-sections along the disclination line show a formation of a zero-intensity region at the axis. 
  (e) Laser beam along the nematic disclination line drawn as isosurface of light intensity (green) at $I/I_0=0.15$; the two headed arrow denotes the polarization of incident light, vertical black lines correspond to the locations of cross-section above. 
  (f-o) Polarizations profiles of light after traversing a distance $z$ along the disclination, starting with linearly polarized light.
  Periodically exchanging patterns of linear and radial polarization are observed.
  Intermediate states (g,i) feature areas with elliptical or circular polarization (shown in yellow and green, respectively). }
 \label{fig:continuous-12}
\end{figure*}

Nematic disclination lines of various winding numbers are used to transform the polarization profile of an incident linearly polarized light beam. 
Figure~\ref{fig:continuous-12} shows the transformation of a linearly polarized light beam travelling along a $s=+1/2$ disclination line.
The incident beam has Gaussian intensity profile (Figure~\ref{fig:continuous-12}a) and linear uniform polarization. 
A zero-intensity region quickly forms at the defect core, and periodic wrinkles are seen in the intensity profile further along the line (Figure~\ref{fig:continuous-12}b-d,e). 
A slight shift of intensity in the $x$ direction is visible, caused by refraction towards the area where the nematic director and light polarization are parallel and the refractive index is higher. 
Intensity profiles at distinct cross-sections also show multiple rings, indicating a presence of higher-order Laguerre-Gaussian modes in the transformed beam. 
Changes in the polarization of light withing the beam are also apparent. 
Interestingly, the incident beam with linear polarization develops a polarization defect at the center, while polarization around the core alternates between two distinct patterns, i.e. the linearly polarized  (Figure~\ref{fig:continuous-12}f,j,l,n) and the radially polarized profile (Figure~\ref{fig:continuous-12}h,k,m,o). 
Note that at each point in space, the polarization of light is linear (as opposed to circular), but the direction of the linear polarization is radial with respect to the center of the incident beam.

\begin{figure*}[!ht]
 \includegraphics[width=\textwidth]{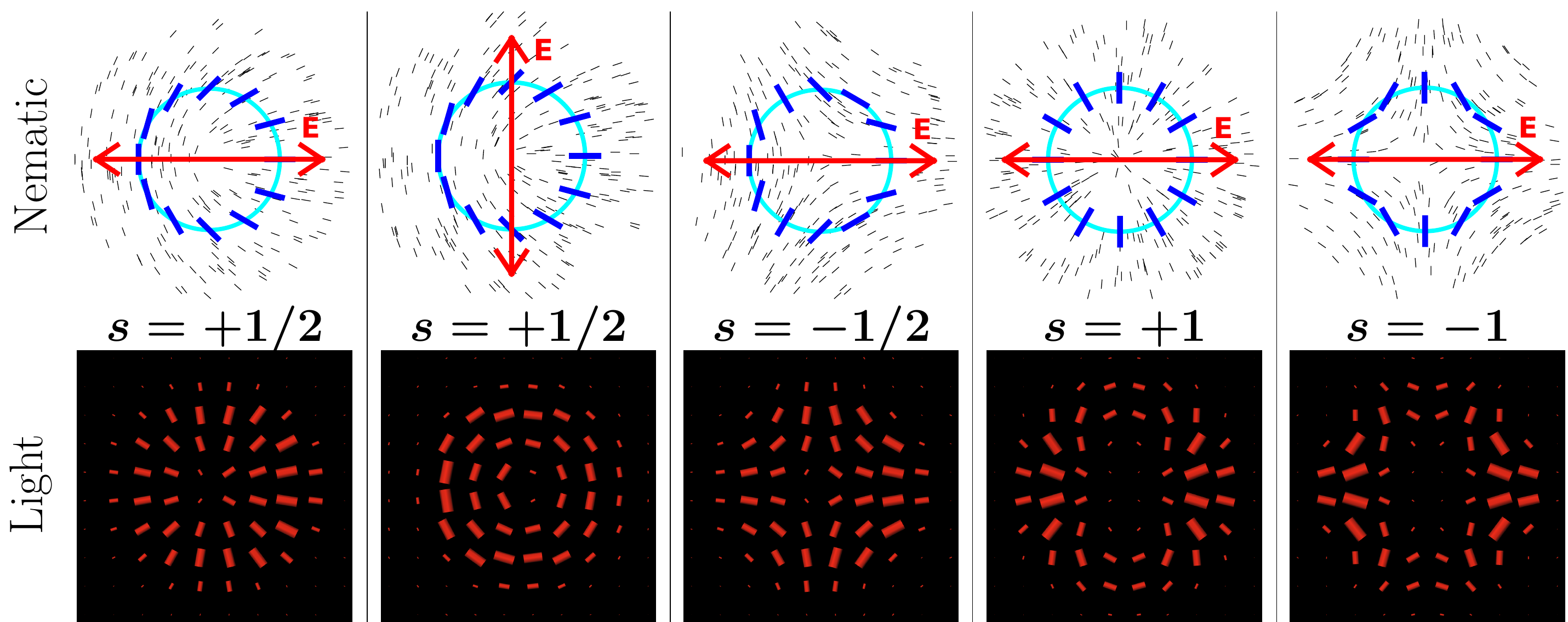}
 \caption{(Color online)
 Generation of complex vector light beams with arbitrary winding numbers via nematic disclinations. The top row shows schematics of nematic disclinations lines with different winding numbers (director in black and dark blue lines) and the (initial) incident linear polarization (red two headed arrows). Corresponding light polarization profiles generated by traversing nematic disclinations and with the initial linear polarization shown above are displayed in the bottom row. All these profiles appear after traversing along the nematic discliantion a distance of $\zd = \lambda / 2\Delta n$. 
}
\label{fig:disclinations}
\end{figure*}

We repeat the simulations with disclinations of different winding numbers and observe that a nematic disclination line -of appropriate length- with winding number $s^{\mathrm{LC}}$ produces a defect in the light polarization with winding number $s^\mathrm{light} = 2s^{\mathrm{LC}}$, as shown in  Figure~\ref{fig:disclinations}. Similarly as with the $+1/2$ disclination shown in Figure~\ref{fig:continuous-12}, we observe that the incident beam with linear polarization develops a polarization defect at the center, while polarization around the core alternates between two distinct patterns, of either linear or $2s^{\mathrm{LC}}$ polarization. It is important to notice that exactly this $2s^{\mathrm{LC}}$ polarization profile developes periodically as the light passes along the nematic disclination line, with the intermediate states characterized by partially linear (shown in red) and partially elliptical (in green) polarization. 
More generally, this systematic study shows that nematic disclination lines could be used as distinct microobjects for controllable generation of polarization profiles in vector light beams, in good agreement with both theoretical predictions and experimental results~\cite{Marucci:06,brasselet-film}. 

The relation between the winding number of the nematic disclination and the winding number of the generated vector light beam, $s^\mathrm{light} = 2s^{\mathrm{LC}}$, can be qualitatively explained by using the Jones formalism.
A phase plate with a retardation of $\delta$ can be described with a Jones matrix $\vec M_\delta$\cite{Marucci:06}:
\begin{equation}\label{eq:jones-matrix-general}
 \vec M_\delta(\alpha) = \vec R(-\alpha) \cdot \begin{pmatrix} 1 & 0 \\ 0 & e^{i\delta} \end{pmatrix} \cdot \vec R(\alpha)
\end{equation}
where $\vec R(\alpha)$ is the rotation matrix by angle $\alpha$, which corresponds to the local orientation of the optical axis, i.e. the nematic director ($\vec n= \cos \alpha \; \hat{\vec e}_x + \sin \alpha \; \hat{\vec e}_y$). For a distinct disclination line with winding numer $s$, the director angle varies as $\alpha = s\phi + \alpha_0$, where $\alpha_0$ is a constant, which for retardation $\delta = \pi$ (also known as the ``q-plate'') gives the Jones matrix:
\begin{equation}
 \vec M_\pi(\alpha) = \begin{pmatrix}\cos 2\alpha & \sin 2\alpha \\ \sin 2\alpha & -\cos 2\alpha\end{pmatrix}  \;.
\end{equation}
Applying $\vec M$ to a linear $x$-polarized incident beam, we obtain
\begin{equation}
 \vec E_{\mathrm{out}} = \vec{M}_\pi \cdot E_0 \begin{pmatrix}1 \\ 0\end{pmatrix} = E_0 \cdot \begin{pmatrix}\cos (2s\phi+\alpha_0) \\ \sin (2s\phi+\alpha_0)\end{pmatrix}
\end{equation}
which is a now a {\it defect light field} with a winding number of $2s$ generated by the defect field of nematic disclination with winding number $s$. 
The distance-of-light-travelled at which the double-winding-number defect emerges  in the light field $z$ is determined by the condition $(2m+1)\pi = \delta = (k_\mathrm{e} - k_\mathrm{o}) z$, from which follows
\begin{align} 
  z = \frac{(2m+1)\lambda}{2(n_\mathrm{e}-n_\mathrm{o})} &= (2m+1)\frac{\lambda}{2\Delta n}, \qquad m\in\mathbb{Z} \nonumber 
\end{align}
where $\lambda$ is the wavelength in vacuum and $\Delta n = n_\mathrm{e} - n_\mathrm{o}$ is the material birefringence. 
At typical values for {\small 5CB} and visible light (see Table \ref{tab:parameters}), the first complete defect in the polarization profile ($m=0$) emerges at
\begin{align}
 \zd = \frac{\SI{480}{\nano\meter}}{2\times 0.15} \approx \SI{1.6}{\micro\meter} \;.
\end{align}
We should stress that the above derivation is only qualitative, as it is based on simple $2\times 2$ Jones calculus which fully neglects refraction. 
For example, in the Jones view, exactly at the disclination, the director angle $\alpha$ is discontinuous, and so is the predicted electric field, which is a non-physical soultion in reality effectively solved by refraction. 
Indeed, our full FDTD numerical simulations clearly show that light refracts away from the defect, creating a zero-intensity region at the disclination.
It is apparent that only methods which treat the electric and magnetic field as full vector fields are suitable for the study of defects in light.
Away from the defect core (more than half a wavelength from the axis), the simplified Jones derivation closely matches our numerical results, and so does the predicted value of $\zd$. 

\subsection{Circular incident polarization}

\begin{figure}[htb]
 \includegraphics[width=\textwidth]{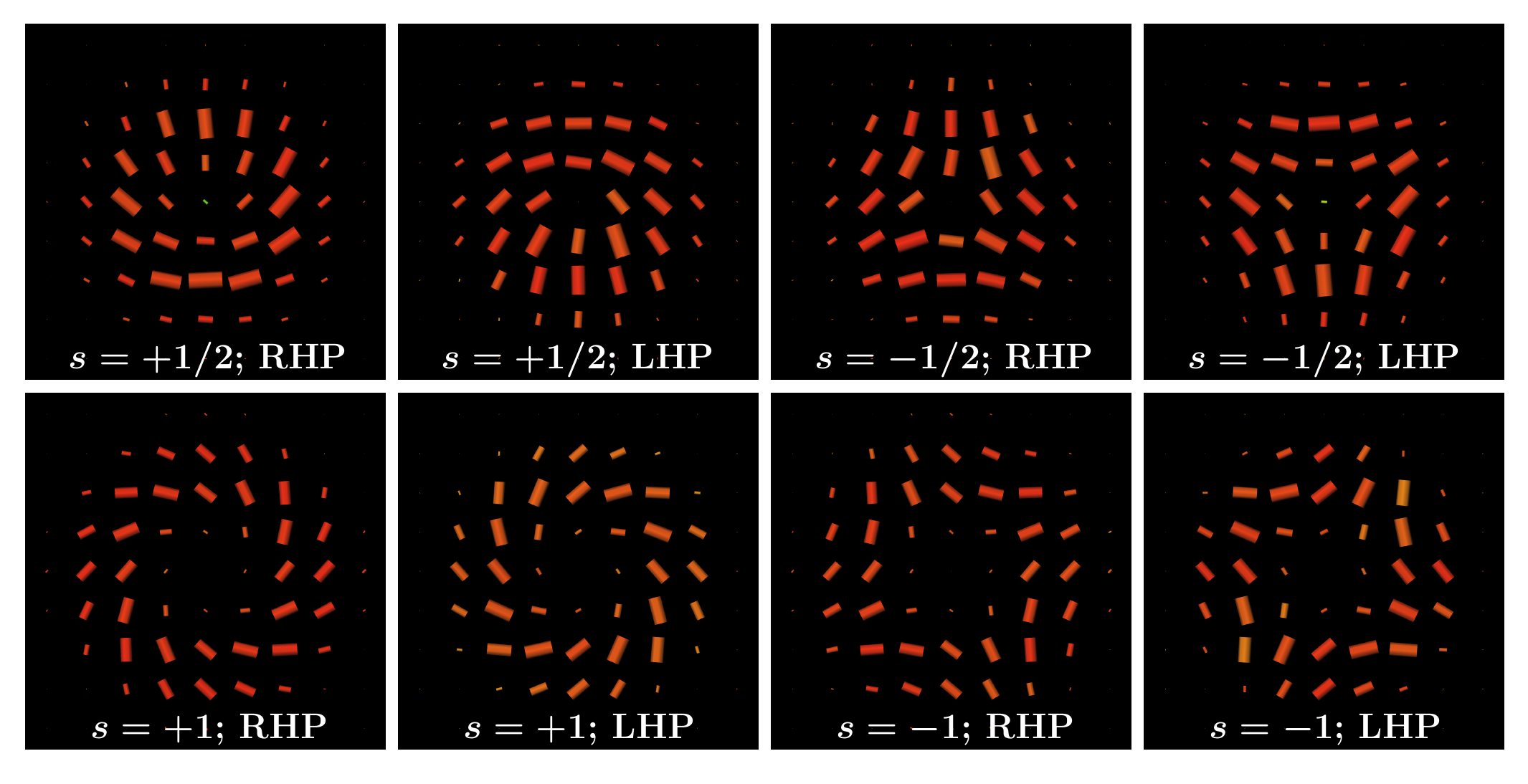}
 \caption{(Color online) Polarization profiles generated from incident circularly-polarized light beams traversing along nematic disclination lines with different winding numbers.  Polarization profiles show defects with the same winding number as the liquid crystal disclination line, notably, including also defects with half-integer winding numbers.
}
 \label{fig:circular}
\end{figure}

By varying the polarization of the incident light beam, it is possible to generate a further large variety of polarization patterns along the nematic disclinations.
Figure \ref{fig:circular} shows polarization profiles of vector light beams obtained by shining incident circularly-polarized light along nematic disclinations.
As in the previous section, a zero-intensity region is always observed at the axis. 
The polarization profiles are shown at distinct propagation lengths, where the polarization again becomes locally linear across the whole cross-section of the beam.
In this case, the defects in the polarization profile appear with the same winding number as that of the disclination line. 
Notably, we observe polarization defects with half-integer winding numbers. 
This is significant because vector fields, such as the electric or magnetic field, can only form defects with integer winding numbers. 
The observed defects thus seemingly violate the topological contraints enforced by the vector symmetry of the light fields. 
However, in the case of light polarization, there are multiple possibilities of rectifying this apparent contradiction \cite{Dennis:Review}. 
At the defect core, there can be an area where light is unpolarized, such as in the polarization pattern of sunlight. 
Alternatively, a point of circularly polarized light (also known as a ``C point'') could be present at the core \cite{Dennis:02}. 
Finally, as it occurs in our case and is explained later in this section, a half-integer strength phase vortex may be present \cite{Angelsky:02}. 

Figure \ref{fig:circular} also shows that the polarization profile is rotated by 45$^\circ$ with respect to the nematic director, with the direction of this rotation depending on the handedness of the circular polarization. For right-handed circular incident polarization ({\small RHP}), the polarization profile rotates by $+45^\circ$ relative to the nematic disclination, whereas for left-handed incident light ({\small LHP}) the profile is rotated by $-45^\circ$, as seen in Figure \ref{fig:circular}. 
Note that while the local polarization at each point is always rotated by $45^\circ$ with respect to the director, this rotation is equivalent to a global rotation by $90^\circ$ in defects with winding numbers $\pm 1/2$ because of the defect's symmetry. 
This symmetry is different in defects with higher winding numbers; in the case of the $s=-1$ defect, the local rotation is equivalent to a global rotation by $22.5^\circ$, while the polarization profile in a $s=+1$ defect cannot be obtained with only global rotations. 

Unlike in the case of linear incident polarization, for circularly polarized incident beams the distinct defect structure in the polarization profile appears at $\zd/2$, where the phase retardation is equal to $\delta = \pi/2$. 
Using the Jones formalism from Eq. \ref{eq:jones-matrix-general}, we can write
\begin{equation}
 \vec M_{\pi/2} = \vec R(-\alpha) \cdot \begin{pmatrix} 1 & 0 \\ 0 & i \end{pmatrix} \cdot \vec R(\alpha)
\end{equation}
\begin{equation}
 \vec E_\mathrm{out} = \vec M_{\pi/2} \cdot \frac{E_0}{\sqrt{2}} \begin{pmatrix}1 \\ i\end{pmatrix} = E_0 e^{i\alpha} \begin{pmatrix}\cos(\alpha -\pi/4) \\ \sin(\alpha - \pi/4)\end{pmatrix}
\end{equation}
Inserting $\alpha = s\phi+\alpha_0$, we indeed obtain a defect in the polarization profile with winding number $s$ that is rotated by $\pi/4$ with respect to the director profile of the liquid crystal disclination line, exactly as found in full {\small FDTD} numerical simulations. 
There are both polarization and phase singularities at the axis, meaning that there the light intensity must drop to zero, which we clearly observe in numerical simulations but is again not accounted for in the Jones formalism. 
As in the case with linear incident polarization, the pattern is repeating with a period of $2\zd$, so these defects can be seen at $z=\zd/2 + 2m\zd$ for each integer $m$. 

\begin{figure}[htb]
\centering
\includegraphics[width=.48\textwidth]{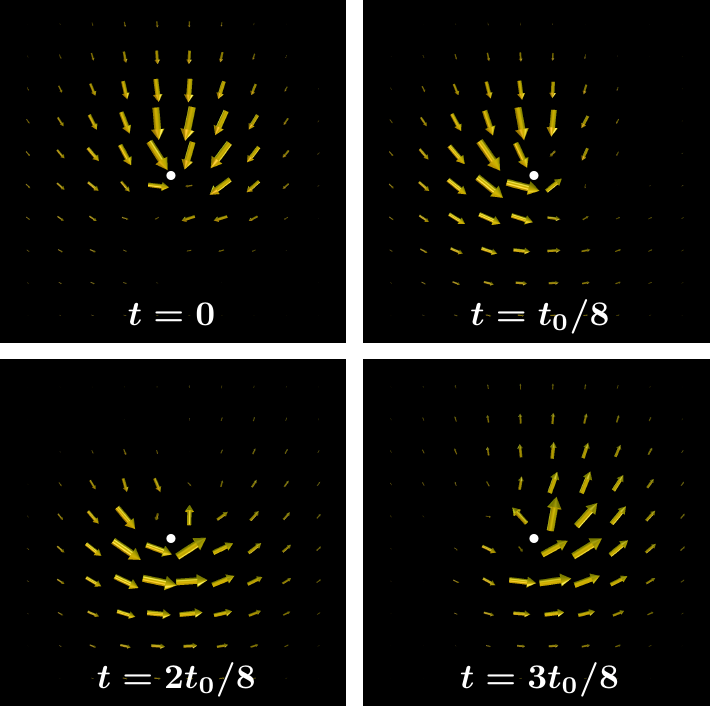}
 \caption{(Color online) Snapshots of electric field in a beam generated from incident circularly-polarized light beam traversing along a nematic disclination line with a different number of $s=+1/2$. 
 The images are taken at different times over half a wave period ($t_0=2\pi/\omega$), demonstrating the phase vortex at the axis (marked by a white dot). 
 }
\label{fig:circular-phase}
\end{figure}

Interestingly, we observe polarization defects with half-integer winding numbers which cannot be formed by electric or magnetic fields and seemingly violate the vector symmetry of the light polarization. 
For a $s=+1/2$ disclination line with $\alpha_0 = 0$, we obtain $\alpha = \phi/2$, resulting in the light field of the form
\begin{equation}
 \vec E_\mathrm{out} = E_0 e^{i\phi/2} \begin{pmatrix}\cos(\phi/2 -\pi/4) \\ \sin(\phi/2- \pi/4)\end{pmatrix} = E_0 e^{i\phi/2} \vec{J}(\phi)
\end{equation}
Considering only the polarization term $\vec J$, we see that $\vec J(2\pi) = -\vec J(0)$, even though the two angles describe the same physical point in space. 
The topological constraints are preserved by a phase vortex at the axis, represented by the term $e^{i\phi/2}$, which can be seen from electric field snapshots in Figure \ref{fig:circular-phase}. 
Along a closed loop around an $s=+1/2$ disclination line, we see a $\pi$ rotation of the polarization and a $\pi$ phase difference, combining into a total $2\pi$ rotation consistent with the vector nature of the electric field.
We thus observe a conversion from a vortexless beam with uniform circular polarization into a beam with a polarization defect and a phase vortex. 
This result clearly demonstrates that the nematic disclination line couples in a complex manner the light's spin, polarization and orbital angular momentum. 

\subsection{Femtosecond pulses along nematic disclinations}

\begin{figure}[h]
 \includegraphics[width=.68\textwidth]{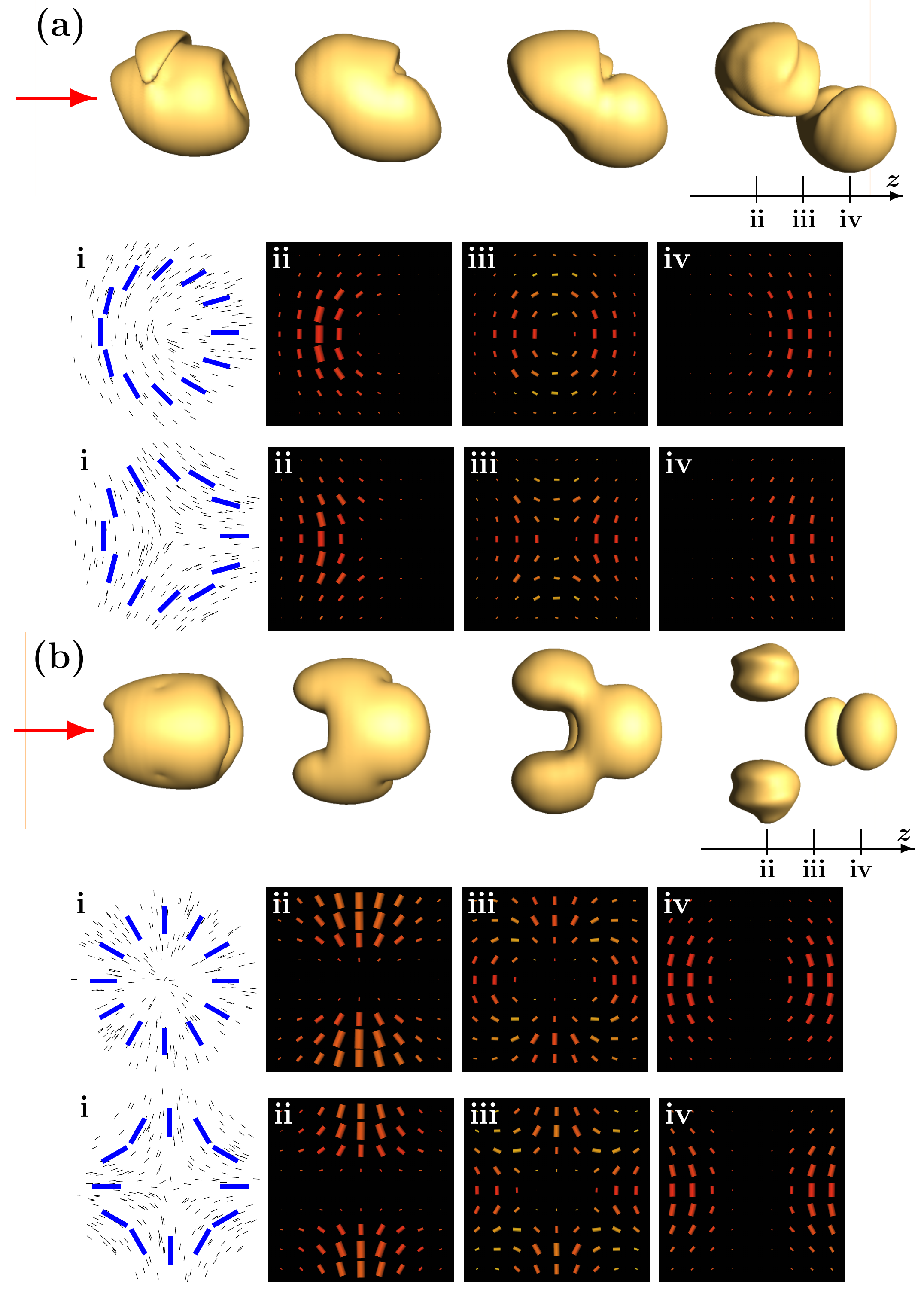}
 \caption{(Color online)
 Intensity regions of a short laser pulse travelling along a nematic disclination line with a winding number (a) $s = \pm 1/2$ and (b) $s = \pm 1$ at four different times. 
 For each winding number, images show the director field (i) and light polarization profile in the last snapshot at three distinct cross sections: the two eigenmodes (ii, iv) and a plane between the main intensity regions (iii). The last snapshot and all cross sections are positioned around \SI{15}{\micro\metre} into the disclination. 
}
 \label{fig:pulse}
\end{figure}

Nematic disclinations can be used also as interesting micro-objects for micro-modulation of laser light {\it pulses}, splitting the pulses into multiple eigenmodes and shaping the pulse intensity and polarization.
Indeed, sending short pulses of light through samples is a strong apporach for finding waveguide modes within the {\small FDTD} method \cite{taflove}.  

In a birefringent medium without a surrounding waveguide, such as the disclination lines we study here, the pulse is a sum of two polarization modes. 
Each of the two modes travels with a different propagation (group) velocity, interestingly, causing the pulse to split into two separate intensity regions with an intensity minimum between them, as shown in Figure \ref{fig:pulse}. 
Additionally, each mode may have multiple intensity regions, depending on the winding number of the nematic disclination. 
An incident femtosecond pulse travelling along a nematic disclination with winding number $s$ gradually splits into two eigenmodes, each of which further divides into $2|s|$ intensity regions, resulting in a total of $4|s|$ intensity regions.
For the $\pm 1/2$ nematic disclinations, we observe effective spliting of the light pulse into 2 intensity regions, whereas for the $\pm 1$ nematic disclinations the pulse splits into 4 intensity regions. 

Our simulations show that while the sign of the winding number is important for the polarization modes, it has no effect on the intensity profile. 
Both eigen polarizations have the same winding number as the director field; however the vector nature of the electric field is incompatible with non-integer defect lines. 
This incompatibility is solved as the director symmetry is broken by the incident polarization. 
Wherever an eigenmode would dictate that the polarization be perpendicular to the incident, light intensity is zero, as this mode is not present in the incident beam.
Topologically more strictly, the polarization of light thus only partially forms topological defect structures, with the light field intensity dropping to zero where the polarization cannot be defined continuously. 
In the region between both propagation modes, the light intensity is relatively low, but interesting polarization features are still visible (Figure \ref{fig:pulse}, profiles $iv$). 
Here, both eigenmodes combine to form a single defect with double winding number and no azimuthal  intensity dependence.
This defect structure is a combination of both polarization modes and is very similar, both in shape and in origin, to those that appear in continuous light. 

\section{Conclusions}

Nematic liquid crystals disclination lines are shown to act as micro-objects which transform the polarization of light and create topological defects in the light field. 
Complex vector light beams of various beam strength are generated from simple linearly and circularly polarized light fields. 
Using numerical modelling based on the FDTD approach, we show that the polarization of light obtains a defect with twice the winding number of the disclination line after traversing a distinct-length section of the line. 
Our results extend the prediction of the Jones method by not only predicting the winding number of light defects but also the corresponding intensity profiles. 
For example, a linearly polarized light beam travelling along a $+1/2$ disclination line can become radially polarized. 
Notably, we show that it is possible to induce polarization defects with half-integer winding numbers by using circularly polarized incident light. 
The topological constraint of the electric field is preserved by a phase vortex, coupling polarization handedness, polarization defects and orbital angular momentum of the beam. 
We further demonstrate that guiding a short laser pulse along a disclination causes it to split into multiple intensity regions. 
These intensity regions are arranged into two propagation eigenmodes, each of which is further divided into $2|s|$ regions, where $s$ is the disclination line's winding number. 

The phenomena described here are interesting for modulation of light polarization, phase and intensity, as well as for spin-orbital momentum transfer, creating light beams with various polarization and phase profiles. 
Due to the inherent susceptibility of liquid crystals to external stimuli, such devices could be further tuned with electric, magnetic or optical fields, offering interesting applications in information processing, possibly as electronic-photonic couplers and as parts of all-photonic circuits. 
Finally, the calculated polarization and intensity profiles will enable us to model the mutual interaction between liquid crystals and light fields in the future, giving a more complete model of light propagation in soft matter. 

\begin{acknowledgments}
The authors acknowledge funding from Slovenian Research Agency Grant Z1-5441 and Programme P1-0099. M.R. acknowledges support from EU FP7 Marie Curie Career Integration Grant FREEFLUID Channel-free liquid crystal microfluidics. 
\end{acknowledgments}


\begin{thebibliography}{46}%
\makeatletter
\providecommand \@ifxundefined [1]{%
 \@ifx{#1\undefined}
}%
\providecommand \@ifnum [1]{%
 \ifnum #1\expandafter \@firstoftwo
 \else \expandafter \@secondoftwo
 \fi
}%
\providecommand \@ifx [1]{%
 \ifx #1\expandafter \@firstoftwo
 \else \expandafter \@secondoftwo
 \fi
}%
\providecommand \natexlab [1]{#1}%
\providecommand \enquote  [1]{``#1''}%
\providecommand \bibnamefont  [1]{#1}%
\providecommand \bibfnamefont [1]{#1}%
\providecommand \citenamefont [1]{#1}%
\providecommand \href@noop [0]{\@secondoftwo}%
\providecommand \href [0]{\begingroup \@sanitize@url \@href}%
\providecommand \@href[1]{\@@startlink{#1}\@@href}%
\providecommand \@@href[1]{\endgroup#1\@@endlink}%
\providecommand \@sanitize@url [0]{\catcode `\\12\catcode `\$12\catcode
  `\&12\catcode `\#12\catcode `\^12\catcode `\_12\catcode `\%12\relax}%
\providecommand \@@startlink[1]{}%
\providecommand \@@endlink[0]{}%
\providecommand \url  [0]{\begingroup\@sanitize@url \@url }%
\providecommand \@url [1]{\endgroup\@href {#1}{\urlprefix }}%
\providecommand \urlprefix  [0]{URL }%
\providecommand \Eprint [0]{\href }%
\providecommand \doibase [0]{http://dx.doi.org/}%
\providecommand \selectlanguage [0]{\@gobble}%
\providecommand \bibinfo  [0]{\@secondoftwo}%
\providecommand \bibfield  [0]{\@secondoftwo}%
\providecommand \translation [1]{[#1]}%
\providecommand \BibitemOpen [0]{}%
\providecommand \bibitemStop [0]{}%
\providecommand \bibitemNoStop [0]{.\EOS\space}%
\providecommand \EOS [0]{\spacefactor3000\relax}%
\providecommand \BibitemShut  [1]{\csname bibitem#1\endcsname}%
\let\auto@bib@innerbib\@empty
\bibitem [{\citenamefont {Hall}(1996)}]{Hall:96}%
  \BibitemOpen
  \bibfield  {author} {\bibinfo {author} {\bibfnamefont {D.~G.}\ \bibnamefont
  {Hall}},\ }\href {\doibase 10.1364/OL.21.000009} {\bibfield  {journal}
  {\bibinfo  {journal} {Opt. Lett.}\ }\textbf {\bibinfo {volume} {21}},\
  \bibinfo {pages} {9} (\bibinfo {year} {1996})}\BibitemShut {NoStop}%
\bibitem [{\citenamefont {Zhan}(2004{\natexlab{a}})}]{Zhan:04}%
  \BibitemOpen
  \bibfield  {author} {\bibinfo {author} {\bibfnamefont {Q.}~\bibnamefont
  {Zhan}},\ }\href {\doibase 10.1364/OPEX.12.003377} {\bibfield  {journal}
  {\bibinfo  {journal} {Opt. Express}\ }\textbf {\bibinfo {volume} {12}},\
  \bibinfo {pages} {3377} (\bibinfo {year} {2004}{\natexlab{a}})}\BibitemShut
  {NoStop}%
\bibitem [{\citenamefont {Tovar}(1998)}]{Tovar:98}%
  \BibitemOpen
  \bibfield  {author} {\bibinfo {author} {\bibfnamefont {A.~A.}\ \bibnamefont
  {Tovar}},\ }\href {\doibase 10.1364/JOSAA.15.002705} {\bibfield  {journal}
  {\bibinfo  {journal} {J. Opt. Soc. Am. A}\ }\textbf {\bibinfo {volume}
  {15}},\ \bibinfo {pages} {2705} (\bibinfo {year} {1998})}\BibitemShut
  {NoStop}%
\bibitem [{\citenamefont {Dorn}\ \emph {et~al.}(2003)\citenamefont {Dorn},
  \citenamefont {Quabis},\ and\ \citenamefont {Leuchs}}]{radial-focus}%
  \BibitemOpen
  \bibfield  {author} {\bibinfo {author} {\bibfnamefont {R.}~\bibnamefont
  {Dorn}}, \bibinfo {author} {\bibfnamefont {S.}~\bibnamefont {Quabis}}, \ and\
  \bibinfo {author} {\bibfnamefont {G.}~\bibnamefont {Leuchs}},\ }\href
  {\doibase 10.1103/PhysRevLett.91.233901} {\bibfield  {journal} {\bibinfo
  {journal} {Phys. Rev. Lett.}\ }\textbf {\bibinfo {volume} {91}},\ \bibinfo
  {pages} {233901} (\bibinfo {year} {2003})}\BibitemShut {NoStop}%
\bibitem [{\citenamefont {Zhan}(2004{\natexlab{b}})}]{Zhan:radial-trap}%
  \BibitemOpen
  \bibfield  {author} {\bibinfo {author} {\bibfnamefont {Q.}~\bibnamefont
  {Zhan}},\ }\href {\doibase 10.1364/OPEX.12.003377} {\bibfield  {journal}
  {\bibinfo  {journal} {Opt. Express}\ }\textbf {\bibinfo {volume} {12}},\
  \bibinfo {pages} {3377} (\bibinfo {year} {2004}{\natexlab{b}})}\BibitemShut
  {NoStop}%
\bibitem [{\citenamefont {Kawauchi}\ \emph {et~al.}(2007)\citenamefont
  {Kawauchi}, \citenamefont {Yonezawa}, \citenamefont {Kozawa},\ and\
  \citenamefont {Sato}}]{radial-trapping-forces}%
  \BibitemOpen
  \bibfield  {author} {\bibinfo {author} {\bibfnamefont {H.}~\bibnamefont
  {Kawauchi}}, \bibinfo {author} {\bibfnamefont {K.}~\bibnamefont {Yonezawa}},
  \bibinfo {author} {\bibfnamefont {Y.}~\bibnamefont {Kozawa}}, \ and\ \bibinfo
  {author} {\bibfnamefont {S.}~\bibnamefont {Sato}},\ }\href {\doibase
  10.1364/OL.32.001839} {\bibfield  {journal} {\bibinfo  {journal} {{Opt.
  Lett.}}\ }\textbf {\bibinfo {volume} {{32}}},\ \bibinfo {pages} {1839}
  (\bibinfo {year} {{2007}})}\BibitemShut {NoStop}%
\bibitem [{\citenamefont {Youngworth}\ and\ \citenamefont
  {Brown}(2000)}]{Youngworth:cvb-focusing-na}%
  \BibitemOpen
  \bibfield  {author} {\bibinfo {author} {\bibfnamefont {K.}~\bibnamefont
  {Youngworth}}\ and\ \bibinfo {author} {\bibfnamefont {T.}~\bibnamefont
  {Brown}},\ }\href {\doibase 10.1364/OE.7.000077} {\bibfield  {journal}
  {\bibinfo  {journal} {Opt. Express}\ }\textbf {\bibinfo {volume} {7}},\
  \bibinfo {pages} {77} (\bibinfo {year} {2000})}\BibitemShut {NoStop}%
\bibitem [{\citenamefont {Niziev}\ and\ \citenamefont
  {Nesterov}(1999)}]{radial-cutting}%
  \BibitemOpen
  \bibfield  {author} {\bibinfo {author} {\bibfnamefont {V.~G.}\ \bibnamefont
  {Niziev}}\ and\ \bibinfo {author} {\bibfnamefont {A.~V.}\ \bibnamefont
  {Nesterov}},\ }\href {http://stacks.iop.org/0022-3727/32/i=13/a=304}
  {\bibfield  {journal} {\bibinfo  {journal} {J. Phys. D Appl. Phys.}\ }\textbf
  {\bibinfo {volume} {32}},\ \bibinfo {pages} {1455} (\bibinfo {year}
  {1999})}\BibitemShut {NoStop}%
\bibitem [{\citenamefont {Kozawa}\ and\ \citenamefont
  {Sato}(2005)}]{kozawa-sato-laser}%
  \BibitemOpen
  \bibfield  {author} {\bibinfo {author} {\bibfnamefont {Y.}~\bibnamefont
  {Kozawa}}\ and\ \bibinfo {author} {\bibfnamefont {S.}~\bibnamefont {Sato}},\
  }\href {\doibase 10.1364/OL.30.003063} {\bibfield  {journal} {\bibinfo
  {journal} {{Opt. Lett.}}\ }\textbf {\bibinfo {volume} {{30}}},\ \bibinfo
  {pages} {3063} (\bibinfo {year} {{2005}})}\BibitemShut {NoStop}%
\bibitem [{\citenamefont {Volpe}\ and\ \citenamefont
  {Petrov}(2004)}]{cvb-fibers}%
  \BibitemOpen
  \bibfield  {author} {\bibinfo {author} {\bibfnamefont {G.}~\bibnamefont
  {Volpe}}\ and\ \bibinfo {author} {\bibfnamefont {D.}~\bibnamefont {Petrov}},\
  }\href {\doibase 10.1016/j.optcom.2004.03.080} {\bibfield  {journal}
  {\bibinfo  {journal} {Opt. Commun.}\ }\textbf {\bibinfo {volume} {237}},\
  \bibinfo {pages} {89–95} (\bibinfo {year} {2004})}\BibitemShut {NoStop}%
\bibitem [{\citenamefont {Shih-Wei~Ko}\ \emph {et~al.}(2010)\citenamefont
  {Shih-Wei~Ko}, \citenamefont {Ting}, \citenamefont {Fuh},\ and\ \citenamefont
  {Lin}}]{polarization-converters-axial}%
  \BibitemOpen
  \bibfield  {author} {\bibinfo {author} {\bibfnamefont {S.-W.}\ \bibnamefont
  {Shih-Wei~Ko}}, \bibinfo {author} {\bibfnamefont {C.-L.}\ \bibnamefont
  {Ting}}, \bibinfo {author} {\bibfnamefont {A.~Y.-G.}\ \bibnamefont {Fuh}}, \
  and\ \bibinfo {author} {\bibfnamefont {T.-H.}\ \bibnamefont {Lin}},\ }\href
  {\doibase 10.1364/OE.18.003601} {\bibfield  {journal} {\bibinfo  {journal}
  {Opt. Express}\ }\textbf {\bibinfo {volume} {18}},\ \bibinfo {pages} {3601}
  (\bibinfo {year} {2010})}\BibitemShut {NoStop}%
\bibitem [{\citenamefont {de~Gennes}\ and\ \citenamefont
  {Prost}(1995)}]{degennes}%
  \BibitemOpen
  \bibfield  {author} {\bibinfo {author} {\bibfnamefont {P.~G.}\ \bibnamefont
  {de~Gennes}}\ and\ \bibinfo {author} {\bibfnamefont {J.}~\bibnamefont
  {Prost}},\ }\href {http://books.google.si/books?id=FDfnnQEACAAJ} {\emph
  {\bibinfo {title} {The Physics of Liquid Crystals, Second Edition}}}\
  (\bibinfo  {publisher} {Oxford University Press},\ \bibinfo {year}
  {1995})\BibitemShut {NoStop}%
\bibitem [{\citenamefont {Kleman}\ and\ \citenamefont
  {Lavrentovich}(2003)}]{kleman}%
  \BibitemOpen
  \bibfield  {author} {\bibinfo {author} {\bibfnamefont {D.}~\bibnamefont
  {Kleman}}\ and\ \bibinfo {author} {\bibfnamefont {O.~D.}\ \bibnamefont
  {Lavrentovich}},\ }\href {http://books.google.si/books?id=dKU9CkazT84C}
  {\emph {\bibinfo {title} {Soft Matter Physics: An Introduction}}}\ (\bibinfo
  {publisher} {Springer},\ \bibinfo {year} {2003})\BibitemShut {NoStop}%
\bibitem [{\citenamefont {Coles}\ and\ \citenamefont
  {Morris}(2010)}]{coles-morris}%
  \BibitemOpen
  \bibfield  {author} {\bibinfo {author} {\bibfnamefont {H.}~\bibnamefont
  {Coles}}\ and\ \bibinfo {author} {\bibfnamefont {S.}~\bibnamefont {Morris}},\
  }\href
  {http://www.nature.com/nphoton/journal/v4/n10/full/nphoton.2010.184.html}
  {\bibfield  {journal} {\bibinfo  {journal} {Nat. Photonics}\ }\textbf
  {\bibinfo {volume} {4}},\ \bibinfo {pages} {676} (\bibinfo {year}
  {2010})}\BibitemShut {NoStop}%
\bibitem [{\citenamefont {Peccianti}\ \emph {et~al.}(2002)\citenamefont
  {Peccianti}, \citenamefont {Conti}, \citenamefont {Assanto}, \citenamefont
  {De~Luca},\ and\ \citenamefont {Umeton}}]{Assanto:solitons}%
  \BibitemOpen
  \bibfield  {author} {\bibinfo {author} {\bibfnamefont {M.}~\bibnamefont
  {Peccianti}}, \bibinfo {author} {\bibfnamefont {C.}~\bibnamefont {Conti}},
  \bibinfo {author} {\bibfnamefont {G.}~\bibnamefont {Assanto}}, \bibinfo
  {author} {\bibfnamefont {A.}~\bibnamefont {De~Luca}}, \ and\ \bibinfo
  {author} {\bibfnamefont {C.}~\bibnamefont {Umeton}},\ }\href {\doibase
  http://dx.doi.org/10.1063/1.1519101} {\bibfield  {journal} {\bibinfo
  {journal} {Appl. Phys. Lett.}\ }\textbf {\bibinfo {volume} {81}},\ \bibinfo
  {pages} {3335} (\bibinfo {year} {2002})}\BibitemShut {NoStop}%
\bibitem [{\citenamefont {Alexander}\ \emph {et~al.}(2012)\citenamefont
  {Alexander}, \citenamefont {Chen}, \citenamefont {Matsumoto},\ and\
  \citenamefont {Kamien}}]{colloquium}%
  \BibitemOpen
  \bibfield  {author} {\bibinfo {author} {\bibfnamefont {G.~P.}\ \bibnamefont
  {Alexander}}, \bibinfo {author} {\bibfnamefont {B.~G.-g.}\ \bibnamefont
  {Chen}}, \bibinfo {author} {\bibfnamefont {E.~A.}\ \bibnamefont {Matsumoto}},
  \ and\ \bibinfo {author} {\bibfnamefont {R.~D.}\ \bibnamefont {Kamien}},\
  }\href {\doibase 10.1103/RevModPhys.84.497} {\bibfield  {journal} {\bibinfo
  {journal} {Rev. Mod. Phys.}\ }\textbf {\bibinfo {volume} {84}},\ \bibinfo
  {pages} {497} (\bibinfo {year} {2012})}\BibitemShut {NoStop}%
\bibitem [{\citenamefont {Araki}\ \emph {et~al.}(2011)\citenamefont {Araki},
  \citenamefont {Buscaglia}, \citenamefont {Bellini},\ and\ \citenamefont
  {Tanaka}}]{Araki:confined}%
  \BibitemOpen
  \bibfield  {author} {\bibinfo {author} {\bibfnamefont {T.}~\bibnamefont
  {Araki}}, \bibinfo {author} {\bibfnamefont {M.}~\bibnamefont {Buscaglia}},
  \bibinfo {author} {\bibfnamefont {T.}~\bibnamefont {Bellini}}, \ and\
  \bibinfo {author} {\bibfnamefont {H.}~\bibnamefont {Tanaka}},\ }\href
  {\doibase 10.1038/NMAT2982} {\bibfield  {journal} {\bibinfo  {journal} {Nat.
  Mater.}\ }\textbf {\bibinfo {volume} {10}},\ \bibinfo {pages} {303} (\bibinfo
  {year} {2011})}\BibitemShut {NoStop}%
\bibitem [{\citenamefont {\v{C}opar}\ \emph {et~al.}(2013)\citenamefont
  {\v{C}opar}, \citenamefont {Clark}, \citenamefont {Ravnik},\ and\
  \citenamefont {\v{Z}umer}}]{opali}%
  \BibitemOpen
  \bibfield  {author} {\bibinfo {author} {\bibfnamefont {S.}~\bibnamefont
  {\v{C}opar}}, \bibinfo {author} {\bibfnamefont {N.~A.}\ \bibnamefont
  {Clark}}, \bibinfo {author} {\bibfnamefont {M.}~\bibnamefont {Ravnik}}, \
  and\ \bibinfo {author} {\bibfnamefont {S.}~\bibnamefont {\v{Z}umer}},\ }\href
  {\doibase 10.1039/C3SM50475A} {\bibfield  {journal} {\bibinfo  {journal}
  {Soft Matter}\ }\textbf {\bibinfo {volume} {9}},\ \bibinfo {pages} {8203}
  (\bibinfo {year} {2013})}\BibitemShut {NoStop}%
\bibitem [{\citenamefont {Brasselet}\ \emph {et~al.}(2009)\citenamefont
  {Brasselet}, \citenamefont {Murazawa}, \citenamefont {Misawa},\ and\
  \citenamefont {Juodkazis}}]{brasselet-droplet}%
  \BibitemOpen
  \bibfield  {author} {\bibinfo {author} {\bibfnamefont {E.}~\bibnamefont
  {Brasselet}}, \bibinfo {author} {\bibfnamefont {N.}~\bibnamefont {Murazawa}},
  \bibinfo {author} {\bibfnamefont {H.}~\bibnamefont {Misawa}}, \ and\ \bibinfo
  {author} {\bibfnamefont {S.}~\bibnamefont {Juodkazis}},\ }\href {\doibase
  10.1103/PhysRevLett.103.103903} {\bibfield  {journal} {\bibinfo  {journal}
  {Phys. Rev. Lett.}\ }\textbf {\bibinfo {volume} {103}},\ \bibinfo {pages}
  {103903} (\bibinfo {year} {2009})}\BibitemShut {NoStop}%
\bibitem [{\citenamefont {Brasselet}(2010)}]{brasselet-reordering}%
  \BibitemOpen
  \bibfield  {author} {\bibinfo {author} {\bibfnamefont {E.}~\bibnamefont
  {Brasselet}},\ }\href {http://stacks.iop.org/2040-8986/12/i=12/a=124005}
  {\bibfield  {journal} {\bibinfo  {journal} {J. Opt.}\ }\textbf {\bibinfo
  {volume} {12}},\ \bibinfo {pages} {124005} (\bibinfo {year}
  {2010})}\BibitemShut {NoStop}%
\bibitem [{\citenamefont {Porenta}\ \emph {et~al.}(2012)\citenamefont
  {Porenta}, \citenamefont {Ravnik},\ and\ \citenamefont
  {\v{Z}umer}}]{porenta}%
  \BibitemOpen
  \bibfield  {author} {\bibinfo {author} {\bibfnamefont {T.}~\bibnamefont
  {Porenta}}, \bibinfo {author} {\bibfnamefont {M.}~\bibnamefont {Ravnik}}, \
  and\ \bibinfo {author} {\bibfnamefont {S.}~\bibnamefont {\v{Z}umer}},\ }\href
  {http://pubs.rsc.org/en/content/articlelanding/2012/sm/c1sm06511d#!divAbstract}
  {\bibfield  {journal} {\bibinfo  {journal} {Soft Matter}\ }\textbf {\bibinfo
  {volume} {8}},\ \bibinfo {pages} {1865} (\bibinfo {year} {2012})}\BibitemShut
  {NoStop}%
\bibitem [{\citenamefont {Loussert}\ \emph {et~al.}(2013)\citenamefont
  {Loussert}, \citenamefont {Delabre},\ and\ \citenamefont
  {Brasselet}}]{brasselet-film}%
  \BibitemOpen
  \bibfield  {author} {\bibinfo {author} {\bibfnamefont {C.}~\bibnamefont
  {Loussert}}, \bibinfo {author} {\bibfnamefont {U.}~\bibnamefont {Delabre}}, \
  and\ \bibinfo {author} {\bibfnamefont {E.}~\bibnamefont {Brasselet}},\ }\href
  {\doibase 10.1103/PhysRevLett.111.037802} {\bibfield  {journal} {\bibinfo
  {journal} {Phys. Rev. Lett.}\ }\textbf {\bibinfo {volume} {111}},\ \bibinfo
  {pages} {037802} (\bibinfo {year} {2013})}\BibitemShut {NoStop}%
\bibitem [{\citenamefont {Marrucci}\ \emph {et~al.}(2006)\citenamefont
  {Marrucci}, \citenamefont {Manzo},\ and\ \citenamefont
  {Paparo}}]{Marucci:06}%
  \BibitemOpen
  \bibfield  {author} {\bibinfo {author} {\bibfnamefont {L.}~\bibnamefont
  {Marrucci}}, \bibinfo {author} {\bibfnamefont {C.}~\bibnamefont {Manzo}}, \
  and\ \bibinfo {author} {\bibfnamefont {D.}~\bibnamefont {Paparo}},\ }\href
  {\doibase 10.1103/PhysRevLett.96.163905} {\bibfield  {journal} {\bibinfo
  {journal} {Phys. Rev. Lett.}\ }\textbf {\bibinfo {volume} {96}},\ \bibinfo
  {pages} {163905} (\bibinfo {year} {2006})}\BibitemShut {NoStop}%
\bibitem [{\citenamefont {Cardano}\ \emph {et~al.}(2013)\citenamefont
  {Cardano}, \citenamefont {Karimi}, \citenamefont {Marrucci}, \citenamefont
  {de~Lisio},\ and\ \citenamefont {Santamato}}]{Cardano:13}%
  \BibitemOpen
  \bibfield  {author} {\bibinfo {author} {\bibfnamefont {F.}~\bibnamefont
  {Cardano}}, \bibinfo {author} {\bibfnamefont {E.}~\bibnamefont {Karimi}},
  \bibinfo {author} {\bibfnamefont {L.}~\bibnamefont {Marrucci}}, \bibinfo
  {author} {\bibfnamefont {C.}~\bibnamefont {de~Lisio}}, \ and\ \bibinfo
  {author} {\bibfnamefont {E.}~\bibnamefont {Santamato}},\ }\href {\doibase
  10.1364/OE.21.008815} {\bibfield  {journal} {\bibinfo  {journal} {Opt.
  Express}\ }\textbf {\bibinfo {volume} {21}},\ \bibinfo {pages} {8815}
  (\bibinfo {year} {2013})}\BibitemShut {NoStop}%
\bibitem [{\citenamefont {Nagali}\ \emph {et~al.}(2009)\citenamefont {Nagali},
  \citenamefont {Sciarrino}, \citenamefont {Martini}, \citenamefont
  {Piccirillo}, \citenamefont {Karimi}, \citenamefont {Marrucci},\ and\
  \citenamefont {Santamato}}]{Nagali:09}%
  \BibitemOpen
  \bibfield  {author} {\bibinfo {author} {\bibfnamefont {E.}~\bibnamefont
  {Nagali}}, \bibinfo {author} {\bibfnamefont {F.}~\bibnamefont {Sciarrino}},
  \bibinfo {author} {\bibfnamefont {F.~D.}\ \bibnamefont {Martini}}, \bibinfo
  {author} {\bibfnamefont {B.}~\bibnamefont {Piccirillo}}, \bibinfo {author}
  {\bibfnamefont {E.}~\bibnamefont {Karimi}}, \bibinfo {author} {\bibfnamefont
  {L.}~\bibnamefont {Marrucci}}, \ and\ \bibinfo {author} {\bibfnamefont
  {E.}~\bibnamefont {Santamato}},\ }\href {\doibase 10.1364/OE.17.018745}
  {\bibfield  {journal} {\bibinfo  {journal} {Opt. Express}\ }\textbf {\bibinfo
  {volume} {17}},\ \bibinfo {pages} {18745} (\bibinfo {year}
  {2009})}\BibitemShut {NoStop}%
\bibitem [{\citenamefont {Cardano}\ \emph {et~al.}(2012)\citenamefont
  {Cardano}, \citenamefont {Karimi}, \citenamefont {Slussarenko}, \citenamefont
  {Marrucci}, \citenamefont {de~Lisio},\ and\ \citenamefont
  {Santamato}}]{Cardano:12}%
  \BibitemOpen
  \bibfield  {author} {\bibinfo {author} {\bibfnamefont {F.}~\bibnamefont
  {Cardano}}, \bibinfo {author} {\bibfnamefont {E.}~\bibnamefont {Karimi}},
  \bibinfo {author} {\bibfnamefont {S.}~\bibnamefont {Slussarenko}}, \bibinfo
  {author} {\bibfnamefont {L.}~\bibnamefont {Marrucci}}, \bibinfo {author}
  {\bibfnamefont {C.}~\bibnamefont {de~Lisio}}, \ and\ \bibinfo {author}
  {\bibfnamefont {E.}~\bibnamefont {Santamato}},\ }\href {\doibase
  10.1364/AO.51.0000C1} {\bibfield  {journal} {\bibinfo  {journal} {Appl.
  Opt.}\ }\textbf {\bibinfo {volume} {51}},\ \bibinfo {pages} {C1} (\bibinfo
  {year} {2012})}\BibitemShut {NoStop}%
\bibitem [{\citenamefont {Humblet}(1943)}]{humblet:43}%
  \BibitemOpen
  \bibfield  {author} {\bibinfo {author} {\bibfnamefont {J.}~\bibnamefont
  {Humblet}},\ }\href {\doibase
  http://dx.doi.org/10.1016/S0031-8914(43)90626-3} {\bibfield  {journal}
  {\bibinfo  {journal} {Physica}\ }\textbf {\bibinfo {volume} {10}},\ \bibinfo
  {pages} {585 } (\bibinfo {year} {1943})}\BibitemShut {NoStop}%
\bibitem [{\citenamefont {Zhao}\ \emph {et~al.}(2007)\citenamefont {Zhao},
  \citenamefont {Edgar}, \citenamefont {Jeffries}, \citenamefont {McGloin},\
  and\ \citenamefont {Chiu}}]{Zhao:spin-orbital}%
  \BibitemOpen
  \bibfield  {author} {\bibinfo {author} {\bibfnamefont {Y.}~\bibnamefont
  {Zhao}}, \bibinfo {author} {\bibfnamefont {J.~S.}\ \bibnamefont {Edgar}},
  \bibinfo {author} {\bibfnamefont {G.~D.~M.}\ \bibnamefont {Jeffries}},
  \bibinfo {author} {\bibfnamefont {D.}~\bibnamefont {McGloin}}, \ and\
  \bibinfo {author} {\bibfnamefont {D.~T.}\ \bibnamefont {Chiu}},\ }\href
  {\doibase 10.1103/PhysRevLett.99.073901} {\bibfield  {journal} {\bibinfo
  {journal} {Phys. Rev. Lett.}\ }\textbf {\bibinfo {volume} {99}},\ \bibinfo
  {pages} {073901} (\bibinfo {year} {2007})}\BibitemShut {NoStop}%
\bibitem [{\citenamefont {Karimi}\ \emph {et~al.}(2009)\citenamefont {Karimi},
  \citenamefont {Piccirillo}, \citenamefont {Marrucci},\ and\ \citenamefont
  {Santamato}}]{Karimi:09}%
  \BibitemOpen
  \bibfield  {author} {\bibinfo {author} {\bibfnamefont {E.}~\bibnamefont
  {Karimi}}, \bibinfo {author} {\bibfnamefont {B.}~\bibnamefont {Piccirillo}},
  \bibinfo {author} {\bibfnamefont {L.}~\bibnamefont {Marrucci}}, \ and\
  \bibinfo {author} {\bibfnamefont {E.}~\bibnamefont {Santamato}},\ }\href
  {\doibase 10.1364/OL.34.001225} {\bibfield  {journal} {\bibinfo  {journal}
  {Opt. Lett.}\ }\textbf {\bibinfo {volume} {34}},\ \bibinfo {pages} {1225}
  (\bibinfo {year} {2009})}\BibitemShut {NoStop}%
\bibitem [{\citenamefont {Berreman}(1972)}]{berreman}%
  \BibitemOpen
  \bibfield  {author} {\bibinfo {author} {\bibfnamefont {D.~W.}\ \bibnamefont
  {Berreman}},\ }\href {\doibase 10.1364/JOSA.62.000502} {\bibfield  {journal}
  {\bibinfo  {journal} {J. Opt. Soc. Am.}\ }\textbf {\bibinfo {volume} {62}},\
  \bibinfo {pages} {502} (\bibinfo {year} {1972})}\BibitemShut {NoStop}%
\bibitem [{\citenamefont {Stallinga}(1999)}]{stallinga-berreman}%
  \BibitemOpen
  \bibfield  {author} {\bibinfo {author} {\bibfnamefont {S.}~\bibnamefont
  {Stallinga}},\ }\href {\doibase http://dx.doi.org/10.1063/1.369638}
  {\bibfield  {journal} {\bibinfo  {journal} {J. Appl. Phys.}\ }\textbf
  {\bibinfo {volume} {85}},\ \bibinfo {pages} {3023} (\bibinfo {year}
  {1999})}\BibitemShut {NoStop}%
\bibitem [{\citenamefont {Kriezis}\ and\ \citenamefont
  {Elston}(2000)}]{metode-kriezis}%
  \BibitemOpen
  \bibfield  {author} {\bibinfo {author} {\bibfnamefont {E.~E.}\ \bibnamefont
  {Kriezis}}\ and\ \bibinfo {author} {\bibfnamefont {S.~J.}\ \bibnamefont
  {Elston}},\ }\href {\doibase http://dx.doi.org/10.1016/S0030-4018(00)00595-2}
  {\bibfield  {journal} {\bibinfo  {journal} {Opt. Commun.}\ }\textbf {\bibinfo
  {volume} {177}},\ \bibinfo {pages} {69 } (\bibinfo {year}
  {2000})}\BibitemShut {NoStop}%
\bibitem [{\citenamefont {Hwang}\ and\ \citenamefont {Rey}(2005)}]{hwang-rey}%
  \BibitemOpen
  \bibfield  {author} {\bibinfo {author} {\bibfnamefont {D.~K.}\ \bibnamefont
  {Hwang}}\ and\ \bibinfo {author} {\bibfnamefont {A.~D.}\ \bibnamefont
  {Rey}},\ }\href {\doibase 10.1364/AO.44.004513} {\bibfield  {journal}
  {\bibinfo  {journal} {Appl. Opt.}\ }\textbf {\bibinfo {volume} {44}},\
  \bibinfo {pages} {4513} (\bibinfo {year} {2005})}\BibitemShut {NoStop}%
\bibitem [{\citenamefont {Oskooi}\ \emph {et~al.}(2010)\citenamefont {Oskooi},
  \citenamefont {Roundy}, \citenamefont {Ibanescu}, \citenamefont {Bermel},
  \citenamefont {Joannopoulos},\ and\ \citenamefont {Johnson}}]{Johnson:Meep}%
  \BibitemOpen
  \bibfield  {author} {\bibinfo {author} {\bibfnamefont {A.~F.}\ \bibnamefont
  {Oskooi}}, \bibinfo {author} {\bibfnamefont {D.}~\bibnamefont {Roundy}},
  \bibinfo {author} {\bibfnamefont {M.}~\bibnamefont {Ibanescu}}, \bibinfo
  {author} {\bibfnamefont {P.}~\bibnamefont {Bermel}}, \bibinfo {author}
  {\bibfnamefont {J.}~\bibnamefont {Joannopoulos}}, \ and\ \bibinfo {author}
  {\bibfnamefont {S.~G.}\ \bibnamefont {Johnson}},\ }\href {\doibase
  http://dx.doi.org/10.1016/j.cpc.2009.11.008} {\bibfield  {journal} {\bibinfo
  {journal} {Comp. Phys. Commun.}\ }\textbf {\bibinfo {volume} {181}},\
  \bibinfo {pages} {687 } (\bibinfo {year} {2010})}\BibitemShut {NoStop}%
\bibitem [{\citenamefont {Taflove}\ and\ \citenamefont
  {Hagness}(2005)}]{taflove}%
  \BibitemOpen
  \bibfield  {author} {\bibinfo {author} {\bibfnamefont {A.}~\bibnamefont
  {Taflove}}\ and\ \bibinfo {author} {\bibfnamefont {S.~C.}\ \bibnamefont
  {Hagness}},\ }\href {http://books.google.si/books?id=n2ViQgAACAAJ} {\emph
  {\bibinfo {title} {Computational Electrodynamics: The Finite-Difference
  Time-Domain Method}}}\ (\bibinfo  {publisher} {Artech House},\ \bibinfo
  {year} {2005})\BibitemShut {NoStop}%
\bibitem [{\citenamefont {Beeckman}\ \emph {et~al.}(2009)\citenamefont
  {Beeckman}, \citenamefont {James}, \citenamefont {Fernandez}, \citenamefont
  {De~Cort}, \citenamefont {Vanbrabant},\ and\ \citenamefont
  {Neyts}}]{Neyts:fem}%
  \BibitemOpen
  \bibfield  {author} {\bibinfo {author} {\bibfnamefont {J.}~\bibnamefont
  {Beeckman}}, \bibinfo {author} {\bibfnamefont {R.}~\bibnamefont {James}},
  \bibinfo {author} {\bibfnamefont {F.}~\bibnamefont {Fernandez}}, \bibinfo
  {author} {\bibfnamefont {W.}~\bibnamefont {De~Cort}}, \bibinfo {author}
  {\bibfnamefont {P.~J.~M.}\ \bibnamefont {Vanbrabant}}, \ and\ \bibinfo
  {author} {\bibfnamefont {K.}~\bibnamefont {Neyts}},\ }\href {\doibase
  10.1109/JLT.2009.2016673} {\bibfield  {journal} {\bibinfo  {journal}
  {Lightwave Technology, Journal of}\ }\textbf {\bibinfo {volume} {27}},\
  \bibinfo {pages} {3812} (\bibinfo {year} {2009})}\BibitemShut {NoStop}%
\bibitem [{\citenamefont {Kriezis}\ and\ \citenamefont
  {Elston}(1999)}]{Kriezis99}%
  \BibitemOpen
  \bibfield  {author} {\bibinfo {author} {\bibfnamefont {E.~E.}\ \bibnamefont
  {Kriezis}}\ and\ \bibinfo {author} {\bibfnamefont {S.~J.}\ \bibnamefont
  {Elston}},\ }\href {\doibase 10.1016/S0030-4018(99)00219-9} {\bibfield
  {journal} {\bibinfo  {journal} {Opt. Commun.}\ }\textbf {\bibinfo {volume}
  {165}},\ \bibinfo {pages} {99} (\bibinfo {year} {1999})}\BibitemShut
  {NoStop}%
\bibitem [{\citenamefont {Ogawa}\ \emph {et~al.}(2013)\citenamefont {Ogawa},
  \citenamefont {ichi Fukuda}, \citenamefont {Yoshida},\ and\ \citenamefont
  {Ozaki}}]{Ogawa:13}%
  \BibitemOpen
  \bibfield  {author} {\bibinfo {author} {\bibfnamefont {Y.}~\bibnamefont
  {Ogawa}}, \bibinfo {author} {\bibfnamefont {J.}~\bibnamefont {ichi Fukuda}},
  \bibinfo {author} {\bibfnamefont {H.}~\bibnamefont {Yoshida}}, \ and\
  \bibinfo {author} {\bibfnamefont {M.}~\bibnamefont {Ozaki}},\ }\href
  {\doibase 10.1364/OL.38.003380} {\bibfield  {journal} {\bibinfo  {journal}
  {Opt. Lett.}\ }\textbf {\bibinfo {volume} {38}},\ \bibinfo {pages} {3380}
  (\bibinfo {year} {2013})}\BibitemShut {NoStop}%
\bibitem [{\citenamefont {Matsui}\ \emph {et~al.}(2014)\citenamefont {Matsui},
  \citenamefont {Kitaguchi},\ and\ \citenamefont {Okajima}}]{Matsui:fdtd}%
  \BibitemOpen
  \bibfield  {author} {\bibinfo {author} {\bibfnamefont {T.}~\bibnamefont
  {Matsui}}, \bibinfo {author} {\bibfnamefont {M.}~\bibnamefont {Kitaguchi}}, \
  and\ \bibinfo {author} {\bibfnamefont {A.}~\bibnamefont {Okajima}},\ }\href
  {\doibase 10.1117/12.2038271} {\bibfield  {journal} {\bibinfo  {journal}
  {Proc. SPIE}\ }\textbf {\bibinfo {volume} {8983}} (\bibinfo {year} {2014}),\
  10.1117/12.2038271}\BibitemShut {NoStop}%
\bibitem [{\citenamefont {Yee}(1966)}]{yee}%
  \BibitemOpen
  \bibfield  {author} {\bibinfo {author} {\bibfnamefont {K.}~\bibnamefont
  {Yee}},\ }\href {\doibase 10.1109/TAP.1966.1138693} {\bibfield  {journal}
  {\bibinfo  {journal} {IEEE T. Antenn. Propag.}\ }\textbf {\bibinfo {volume}
  {14}},\ \bibinfo {pages} {302} (\bibinfo {year} {1966})}\BibitemShut
  {NoStop}%
\bibitem [{\citenamefont {Werner}\ and\ \citenamefont
  {Cary}(2007)}]{Werner:2007:stabilnost}%
  \BibitemOpen
  \bibfield  {author} {\bibinfo {author} {\bibfnamefont {G.~R.}\ \bibnamefont
  {Werner}}\ and\ \bibinfo {author} {\bibfnamefont {J.~R.}\ \bibnamefont
  {Cary}},\ }\href {\doibase http://dx.doi.org/10.1016/j.jcp.2007.05.008}
  {\bibfield  {journal} {\bibinfo  {journal} {J. Comput. Phys.}\ }\textbf
  {\bibinfo {volume} {226}},\ \bibinfo {pages} {1085 } (\bibinfo {year}
  {2007})}\BibitemShut {NoStop}%
\bibitem [{\citenamefont {{Berenger}}(1994)}]{berenger}%
  \BibitemOpen
  \bibfield  {author} {\bibinfo {author} {\bibfnamefont {J.-P.}\ \bibnamefont
  {{Berenger}}},\ }\href {\doibase 10.1006/jcph.1994.1159} {\bibfield
  {journal} {\bibinfo  {journal} {J. Comput. Phys.}\ }\textbf {\bibinfo
  {volume} {114}},\ \bibinfo {pages} {185} (\bibinfo {year}
  {1994})}\BibitemShut {NoStop}%
\bibitem [{\citenamefont {Chirtoc}\ \emph {et~al.}(2004)\citenamefont
  {Chirtoc}, \citenamefont {Chirtoc}, \citenamefont {Glorieux},\ and\
  \citenamefont {Thoen}}]{kolicniki}%
  \BibitemOpen
  \bibfield  {author} {\bibinfo {author} {\bibfnamefont {I.}~\bibnamefont
  {Chirtoc}}, \bibinfo {author} {\bibfnamefont {M.}~\bibnamefont {Chirtoc}},
  \bibinfo {author} {\bibfnamefont {C.}~\bibnamefont {Glorieux}}, \ and\
  \bibinfo {author} {\bibfnamefont {J.}~\bibnamefont {Thoen}},\ }\href
  {\doibase 10.1080/02678290310001642540} {\bibfield  {journal} {\bibinfo
  {journal} {Liq. Cryst.}\ }\textbf {\bibinfo {volume} {31}},\ \bibinfo {pages}
  {229} (\bibinfo {year} {2004})}\BibitemShut {NoStop}%
\bibitem [{\citenamefont {Dennis}\ \emph {et~al.}(2009)\citenamefont {Dennis},
  \citenamefont {O'Holleran},\ and\ \citenamefont {Padgett}}]{Dennis:Review}%
  \BibitemOpen
  \bibfield  {author} {\bibinfo {author} {\bibfnamefont {M.~R.}\ \bibnamefont
  {Dennis}}, \bibinfo {author} {\bibfnamefont {K.}~\bibnamefont {O'Holleran}},
  \ and\ \bibinfo {author} {\bibfnamefont {M.~J.}\ \bibnamefont {Padgett}},\
  }\href {\doibase http://dx.doi.org/10.1016/S0079-6638(08)00205-9} {\bibfield
  {journal} {\bibinfo  {journal} {Prog. Opt.}\ }\textbf {\bibinfo {volume}
  {53}},\ \bibinfo {pages} {293} (\bibinfo {year} {2009})}\BibitemShut
  {NoStop}%
\bibitem [{\citenamefont {Dennis}(2002)}]{Dennis:02}%
  \BibitemOpen
  \bibfield  {author} {\bibinfo {author} {\bibfnamefont {M.}~\bibnamefont
  {Dennis}},\ }\href {\doibase http://dx.doi.org/10.1016/S0030-4018(02)02088-6}
  {\bibfield  {journal} {\bibinfo  {journal} {Opt. Commun.}\ }\textbf {\bibinfo
  {volume} {213}},\ \bibinfo {pages} {201} (\bibinfo {year}
  {2002})}\BibitemShut {NoStop}%
\bibitem [{\citenamefont {Angelsky}\ \emph {et~al.}(2002)\citenamefont
  {Angelsky}, \citenamefont {Mokhun}, \citenamefont {Mokhun},\ and\
  \citenamefont {Soskin}}]{Angelsky:02}%
  \BibitemOpen
  \bibfield  {author} {\bibinfo {author} {\bibfnamefont {O.}~\bibnamefont
  {Angelsky}}, \bibinfo {author} {\bibfnamefont {A.}~\bibnamefont {Mokhun}},
  \bibinfo {author} {\bibfnamefont {I.}~\bibnamefont {Mokhun}}, \ and\ \bibinfo
  {author} {\bibfnamefont {M.}~\bibnamefont {Soskin}},\ }\href {\doibase
  http://dx.doi.org/10.1016/S0030-4018(02)01479-7} {\bibfield  {journal}
  {\bibinfo  {journal} {Opt. Commun.}\ }\textbf {\bibinfo {volume} {207}},\
  \bibinfo {pages} {57} (\bibinfo {year} {2002})}\BibitemShut {NoStop}%
\end{thebibliography}
\end{document}